\documentclass[a4paper,superscriptaddress,showkeys,prX,showpacs,twocolumn]{revtex4-1}
\usepackage{amssymb, amsmath, amsfonts}
\usepackage{graphics}
\usepackage{pst-all}
\usepackage{hyperref}
\usepackage[displaymath,mathlines,running]{lineno}
\usepackage{relsize}
\usepackage[utf8]{inputenc}
\usepackage{IEEEtrantools}
\usepackage{epsfig}

\begin{document}
\title{Spin current injection at magnetic insulator/superconductor interfaces}
\author{V. S. U. A. Vargas}
\email{santunionivinicius@gmail.com}
\author{A. R. Moura}
\email{antoniormoura@ufv.br}
\affiliation{Departamento de F\'{i}sica, Universidade Federal de Vi\c{c}osa, 36570-900, Vi\c{c}osa, Minas Gerais, Brazil}
\date{\today}
\begin{abstract}
Opposite to the common idea of a magnetic order requirement to obtain spin current propagation, 
materials with no magnetic ordering have also been revealed to be efficient spin conductors. In this work, we 
investigate the spin current injection at the interface between a magnetic insulator and a 
superconductor. We are mainly interested in the paramagnetic insulator/superconductor interface
however, our model also describes the ferromagnetic phase. We used the Schwinger bosonic formalism 
to describe the magnetic insulator and standard BCS theory was applied to treat the 
superconductor layer. In the normal-metal limit, our results are in agreement with the expected ones. 
For example, we found the correct spin current behavior $I\approx T^{3/2}$ at low temperature. 
In addition, our model shows a pronounced peak in the spin current injection at temperatures 
close to the superconductor transition temperature due to the superconducting
quasiparticle coherence. The role of magnetic fields in the spin current injection is also investigated.
\end{abstract}
  

\maketitle

\section{Introduction and motivation}
Charge currents were the basis of a very large technological development in the 20th 
century. Even today, most commercial devices are fundamentally electronic-based ones. 
However, in recent years, spintronics research has taken pride of place in 
the scientific community. The continuous advance in miniaturization has supported the 
generation, manipulation, and detection of spin current in many different material classes. 
Basically, spin current involves effective spin transport that can be followed or 
not by electrical charge current. In a ferromagnetic conductor, for example, due to the 
electron spin-polarization the current transports spin and charge at the same time. On the other 
hand, pure spin currents can be obtained when charge currents of opposite spins move in 
opposite directions, which occurs in metals with strong spin-orbit interaction, the 
so-called spin Hall effect \cite{PRL83.1834,prl85.393,rmp87.1213,prb94.104419,prb94.104420}. 
In insulators, the spin current is driven by magnons (or spin waves in the classical 
formalism) and is observed in ferromagnetic  \cite{prl88.117601,prb83.144402,jpcs200.062030}, 
antiferromagnetic  \cite{rmp90.015005,prb89.140406,prb90.094408,prb93.054412,prb94.014412,prl116.186601},
 and paramagnetic (PMI) insulators \cite{prl116.186601,naturecomm10.1,naturephys13.987}. 
Temperature gradients (spin Seebeck effect, SSE) \cite{prb83.094410,rpp76.036501} as well as 
time-dependent ferromagnetic magnetization (spin pumping, SP) \cite{prb83.144402,prb89.174417,JAP97.10C715} 
are frequently used to generate spin current in adjacent materials. 
The detection of spin current in conductors can be performed by the inverse spin Hall 
effect (ISHE) \cite{APL88.182509,prl98.156601,nature442.176}, where a transverse 
charge current provides a detectable bias voltage. In addition, when spin current is injected 
into (from) a magnetic insulator, the decrease (increase) in Gilbert damping is detected 
by measurements of the microwave radiation emitted in the ferromagnetic resonance (FMR) \cite{prl88.117601,prb66.224403}.

Although it is usual to consider spin current injection in ferromagnetic materials, 
an ordered state is not really a necessary condition in spintronics. Indeed, Shiomi and 
Saitoh verified SP in the paramagnetic insulator La$_2$NiMnO$_6$ \cite{prl113.266602}, 
while Wu {\it et al.} performed measurements of paramagnetic SSE in DySCO$_3$ and 
Gd$_3$Ga$_5$O$_{12}$ [gadolinium gallium garnet, GGG) \cite{prl114.186602}]. 
A theoretical model to describe SSE in paramagnets and antiferromagnets (both phases 
without a magnetization order) was developed by Yamamoto {\it et al.} \cite{prb100.064419}. 
Curiously, GGG is a well known substrate for growing superconductor films and FM layers 
of yttrium iron garnet (YIG) but only recently has it been directly applied in spin 
transport experiments. Due to the very low exchange coupling $J_\textrm{ex}\approx 100$ mK
 (8.6 $\mu$eV), GGG presents a low Curie temperature transition $T_c\approx$ 180 mK. 
Recently, Oyanagi {\it et al.} demonstrated the efficiency of transporting spin in a GGG slab even at
temperatures several orders above $T_c$ \cite{naturecomm10.1}. Amorphous-YIG is a 
paramagnet that also presents efficient spin transport  \cite{naturephys13.987}. 
Therefore, there is much evidence for the unnecessary condition of magnetic ordering 
in spin current propagation.

In this work we investigate the spin current injection from a superconductor (SC) 
into a paramagnetic insulator. The charge current injection at superconducting interfaces 
has been well known since the early 1980s. Spin-polarized quasiparticles were observed in an
s-wave superconductor due to injected spin-polarized charge current as well as spin 
accumulation and spin diffusion in superconducting samples \cite{PRL55.1790,PRB37.5326,APL65.1460}. 
On the other hand, spin current injection at superconducting interfaces is a more recent topic. 
In Ref.  \cite{prl100.047002}, for example, the authors determined the influence of 
superconductivity in spin current through measurements of the Gilbert damping 
in Ni$_{80}$Fe$_{20}$ films grown on Nb. Yao {\it et al.} also investigated the spin 
dynamics at interfaces composed of superconducting NbN films and the ferromagnetic 
insulator GdN \cite{prb97.224414}. Theoretical models to describe spin current injection
at SC/FM interfaces can be found in Refs.  \cite{prb96.024414,prb99.144411,jmmm494.165813}.
The scenario involving a paramagnetic insulator/normal-metal (PMI/NM) junction was analyzed by 
Okamoto \cite{prb93.064421}. Okamoto used the Schwinger bosonic formalism to determine 
the spin current injected and spin conductivity. Here we also adopt Schwinger 
bosons to describe the disordered phase in terms of spinon operators that interact 
with quasiparticles in the SC through an sd-interaction at the interface. Therefore, the 
developed model is useful for describing the spin current injection in both NM and SC phases at the 
interface with FM or PM insulators. We found results compatible with similar experiments 
and according to the special-limit cases, for example, the superconducting gapless 
$\Delta=0$ phase. It is important to note that our model corrects the discrepancy in 
the temperature dependence of the spin current $I$ at NM/FM interfaces found in Ref. \cite{prb93.064421} 
and provides the expected $I(T)\propto T^{3/2}$ behavior at low temperatures. 
In the PMI/SC junction we found a pronounced peak in 
the spin current injection due to the quasiparticle coherence. In addition, the
spin conductance dependence on external magnetic fields is investigated. We verify a
decreasing spin current with increasing magnetic fields due to the quasi particle
creation restraint.

\section{Model and methods}
The studied model is described by the Hamiltonian $H=H_\textrm{m}+H_\textrm{SC}+H_\textrm{sd}$,
where the terms define the magnetic insulator, the superconductor, and the interface 
interaction, respectively. Both magnetic and superconductor sides are considered 
three-dimensional samples, but a model of thin films can be treated with minor 
modifications. The sd Hamiltonian represents an interaction at the interface 
between located electrons of the insulator and conduction electrons of the normal-metal. 
In this section, we briefly review the main points of the Schwinger formalism to represent magnetic models and the microscopic BCS theory.

The magnetic insulator is given by the standard Heisenberg Hamiltonian 
$H_\textrm{m}=-J\sum_{\langle ij\rangle}\vec{S}_i\cdot\vec{S}_j$, where $J$ 
is a small exchange ferromagnetic coupling and the sum is taken over nearest-neighbor spins. 
At low temperature, spin operators are commonly treated by using the
Holstein-Primakoff (HP) bosonic representation. However, HP bosons are
inaccurate for representing disordered magnetic phases. A more appropriate representation
is obtained through Schwinger bosons, which are applicable to both ordered and disordered
phases  \cite{prb38.316,prb40.5028}. The spin operators are then replaced by two kinds of 
bosonic operators and written as 
$S_i^+=a_{i\uparrow}^\dagger a_{i\downarrow}$, $S_i^-=a_{i\downarrow}^\dagger a_{i\uparrow}$, 
and $S_i^z=(a_{i\uparrow}^\dagger a_{i\uparrow}-a_{i\downarrow}^\dagger a_{i\downarrow})/2$, 
where $a_{i\sigma}^\dagger$ ($a_{i\sigma}$) creates (annihilates) a spinon with spin 
$\sigma/2$ ($\sigma=1$ stands for up spin and $\sigma=-1$ stands for down spin). 
To ensure the commutation relation 
$[S_i^a,S_j^b]=i\delta_{ij}\epsilon_{abc}S_i^c$ is necessary to fix the number 
of bosons on each site through the constraint $\sum_\sigma a_{i\sigma}^\dagger a_{i\sigma}=2S$. The Hamiltonian is then given by

\begin{figure}[h]
\centering \epsfig{file=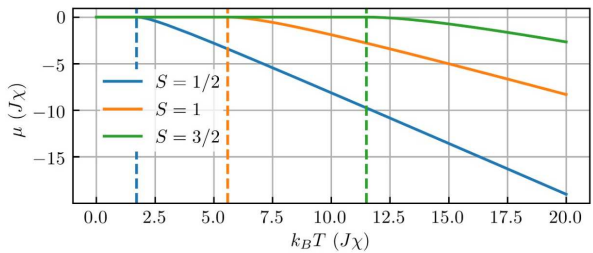,width=0.9\linewidth}
\caption{The chemical potential $\mu_\textrm{m}$ as a function of temperature for $B=0$. The vertical dashed lines represent the
critical temperature of the boson condensation.}
\label{fig.mu}
\end{figure}

\begin{IEEEeqnarray}{rCl}
H_\textrm{m}&=&-\frac{J}{2}\sum_{\langle ij\rangle} (:\mathcal{F}_{ij}^\dagger \mathcal{F}_{ij}:-2S^2)+\sum_i \lambda_i(\mathcal{F}_{ii}-2S)-\nonumber\\
&&-\frac{g\mu_B B}{2}\sum_i(a_{i\uparrow}^\dagger a_{i\uparrow}-a_{i\downarrow}^\dagger a_{i\downarrow})
\end{IEEEeqnarray}
in which we defined the bond operator 
$\mathcal{F}_{ij}=a_{i\uparrow}^\dagger a_{j\uparrow}+a_{i\downarrow}^\dagger a_{j\downarrow}$ 
and $: :$ represents the normal ordering operator. We include a uniform 
magnetic field $\vec{B}=B\hat{z}$, and the constraint is implemented by a local 
Lagrange multiplier $\lambda_i$. The quartic order term is decoupled by introducing 
an auxiliary field $F_{ij}=\langle\mathcal{F}_{ij}\rangle$ through the 
Hubbard-Stratonovich transform 
$\mathcal{F}_{ij}^\dagger \mathcal{F}_{ij}\to F_{ij}(\mathcal{F}_{ij}^\dagger+\mathcal{F}_{ij})-F_{ij}^2$.
We consider a mean-field theory and replace $F_{ij}$ by a uniform field 
$F$. In the same way we approximate the Lagrange multiplier by a uniform parameter 
$\lambda$, which implies boson conservation only on average. After a space
Fourier transform, we obtain the quadratic Hamiltonian

\begin{equation}
H_\textrm{m}=E_0+\sum_q\left[\hbar\Omega_{q\uparrow}a_{q\uparrow}^\dagger a_{q\uparrow}+\hbar\Omega_{q\downarrow} a_{q\downarrow}^\dagger a_{q\downarrow}\right],
\end{equation}
where $E_0=3NJ(F^2+2S^2)/2-2NS(3JF-\mu_\textrm{m})$ is the ground-state energy and 
$\hbar\Omega_{q\sigma}=\hbar\omega_q-\mu_\textrm{m}-\sigma g\mu_B B/2$. 
In the above equation, $N$ is the number of magnetic sites, $\hbar\omega_q=3JF(1-\gamma_q)$, and 
$\gamma_q=(\cos q_x+\cos q_y+\cos q_z)/3$ is the square lattice structure factor. 
The chemical potential $\mu_\textrm{m}=3JF-\lambda$ was introduced to make clear
the analogy between the ordered phase transition and Bose-Einstein condensation 
($-\mu_\textrm{m}$ could also be considered a gap in spectrum energy \cite{auerbach}). 
The fields $F$ and $\lambda$ are evaluated by the minimization of the free energy 
$F_\textrm{m}=-\beta^{-1}\ln(\textrm{Tr} e^{-\beta H_\textrm{m}})$. 
The extremum conditions $\delta F_\textrm{m}/\delta\lambda=0$ and 
$\delta F_\textrm{m}/\delta F=0$ provide the self-consistent equations

\begin{equation}
\label{eq.sce1}
2S=\frac{1}{N}\sum_q(n_{q\uparrow}+n_{q\downarrow})
\end{equation}
and
\begin{equation}
\label{eq.sce2}
F=2S-\frac{1}{N}\sum_q\frac{\hbar\omega_q}{3JF}(n_{q\uparrow}+n_{q\downarrow}),
\end{equation}
where $n_{q\sigma}=(e^{\beta\hbar\Omega_{q\sigma}}-1)^{-1}$ 
is the Bose-Einstein distribution. In the disordered 
phase $\mu_\textrm{m}+g\mu_B B/2<0$, and the self-consistent equations present a non-trivial solution 
for $F$ and $\lambda$. At a critical temperature we obtain $\mu_\textrm{m}=-g\mu_B B/2$ and the 
up-spin boson condensate in the $q=0$ state ($\Omega_{0\uparrow}=0)$. As well as in standard Bose-Einstein 
condensation, we separate the $q=0$ term from the sum before converting it to a momentum integral to solve 
the equations in the ordered phase. At low temperature, the second self-consistent 
equation provides $F\approx 2S$. We then include a phenomenological parameter $\chi$ to 
consider a small correction and express $F=2S\chi$ ($\chi=1$ in the limit $T\to 0$). In 
the long-wavelength limit, $\chi$ can be determined by the deviation magnetization per 
site given by 

\begin{equation}
\frac{\Delta m}{N}= S-\langle S^z\rangle \approx\left(\frac{k_B T}{4\pi J\chi S}\right)^{3/2},
\end{equation}
adopting the limit $B\to 0^+$. Here and henceforth, we will use $J\chi$ as the magnetic energy scale. 
Using Eq. (\ref{eq.sce1}), we then determine the dependence of the chemical potential $\mu_\textrm{m}$
on temperature and magnetic field. A graphic of $\mu_\textrm{m}(T)$ for $B=0$ is shown in Fig. \ref{fig.mu} 
(cases with finite $B$ present similar behavior).

The superconductor is described by the well known BCS theory \cite{tinkham}, whose 
Hamiltonian is written as

\begin{equation}
H_\textrm{sc}=\sum_{k\sigma}\epsilon_k c_{k\sigma}^\dagger c_{k\sigma}-g_\textrm{eff}\sum_{kk^\prime}
c_{k\uparrow}^\dagger c_{-k\downarrow}^\dagger c_{-k^\prime\downarrow}c_{k^\prime\uparrow},
\end{equation} 
in which $g_\textrm{eff}$ is the effective superconducting interaction constant. 
The momentum sum is done within the range $\pm\hbar\omega_\textrm{D}$ of the Fermi surface, 
{\it i.e.}, $|\epsilon_k-\epsilon_\textrm{F}|<\hbar\omega_\textrm{D}$, 
where $\epsilon_\textrm{F}$ is the Fermi energy and $\omega_\textrm{D}$ is the Debye frequency. 
Typical energy scales for the Fermi and Debye energy are 10 and $10^{-2}$ eV, respectively.
A population imbalance between up- and down-spin electrons is necessary to ensure spin
current injection from the SC. After including the chemical potentials $\mu_\uparrow$ 
and $\mu_\downarrow$ for up- and down-spin electrons, respectively, the grand-canonical Hamiltonian 
is expressed as

\begin{equation}
K_\textrm{SC}=C+\sum_k\Psi_k^\dagger\left(\begin{array}{cc} \xi_k-\mu_\textrm{sc} & -\Delta \\
-\bar{\Delta} &-\xi_k-\mu_\textrm{sc}\end{array}\right)\Psi_k
\end{equation}
where $C=|\Delta|^2/g_\textrm{eff}+\sum_k(\xi_k+\mu_\textrm{sc})$ is a 
constant and the Nambu spinor is defined by $\Psi_k^\dagger=(c_{k\uparrow}^\dagger\ \ c_{-k\downarrow})$. 
The quartic order interaction was decoupled by introducing the superconducting gap
$\Delta=g_\textrm{eff}\sum_k\langle c_{-k\downarrow}c_{k\uparrow}\rangle$. In the above equation
$\xi_k=\epsilon_k-(\mu_\uparrow+\mu_\downarrow)/2$, and the SC chemical potential (the Zeeman splitting)
is defined as $\mu_\textrm{sc}=(g\mu_B B+\Delta\mu)/2$, with 
$\Delta\mu=\mu_\uparrow-\mu_\downarrow$. Here we also include the uniform magnetic 
field $\vec{B}=B\hat{z}$. 
While the superconducting ground-state is composed of Cooper pairs, the excitations are given 
by quasiparticles (also called bogoliubons) of energy $E_k=\sqrt{\xi_k^2+|\Delta|^2}$. The BCS Hamiltonian is diagonalized defining new fermionic operators by the Bogoliubov transform

\begin{figure}[h]
\centering \epsfig{file=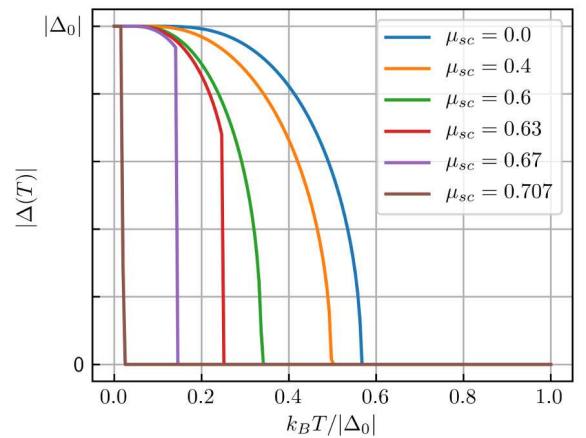,width=0.9\linewidth}
\caption{The gap dependence on temperature for different values of $\mu_\textrm{sc}$. Above the SC 
transition temperature the gap vanishes. For $\mu_\textrm{sc}>0.707|\Delta_0|$ the SC is suppressed 
even at zero temperature.}
\label{fig.gap_T}
\end{figure}

\begin{IEEEeqnarray}{rCl}
\IEEEyesnumber
\IEEEyessubnumber*
\label{eq.b}
b_{k\uparrow}&=&\bar{u}_k c_{k\uparrow}+v_k c_{-k\downarrow}^\dagger\\
b_{k\downarrow}&=&\bar{u}_k c_{k\downarrow}-v_k c_{-k\uparrow}^\dagger,
\end{IEEEeqnarray}
with the parameters $u_k=e^{-i\phi/2}\sqrt{(E_k+\xi_k)/2E_k}$ and $v_k=e^{i\phi/2}\sqrt{(E_k-\xi_k)/2E_k}$ 
($\phi$ is the superconducting gap phase, $\Delta=e^{i\phi}|\Delta|$). The diagonal BCS Hamiltonian
is then given by

\begin{equation}
K_\textrm{SC}=K_0+\sum_k(E_{k\uparrow}b_{k\uparrow}^\dagger b_{k\uparrow}+E_{k\downarrow}b_{k\downarrow}^\dagger b_{k\downarrow}),
\end{equation}
where $K_0=|\Delta|^2/g_\textrm{eff}+\sum_k(\xi_k-E_k)$ is a constant energy and 
$E_{k\sigma}=E_k-\sigma\mu_\textrm{sc}$. Using the above Hamiltonian, we
obtain the self-consistent gap equation
\begin{equation}
\label{eq.gapsc}
\Delta=\sum_k \frac{g\Delta}{4E_k}\left[\tanh\left(\frac{\beta E_{k\uparrow}}{2}\right)+\tanh\left(\frac{\beta E_{k\downarrow}}{2}\right)\right],
\end{equation} 
which provides the result in Fig. \ref{fig.gap_T}. The SC temperature transition is 
defined as the temperature at which the gap vanishes. For $\mu_\textrm{sc}>0.707|\Delta_0|$
the SC is suppressed even at zero temperature.

As one can see, the chemical potential difference between up and down spin quasiparticles 
favors processes with annihilation (creation) of up spin (down spin) quasiparticles. 
Therefore, a positive value of $\mu_\textrm{sc}$ provides a spin current flux from the SC 
into the magnetic insulator.
However, the presence of polarizing terms such as the magnetic field and the chemical potential 
imbalance $\Delta\mu$ tend to destroy the superconducting phase \cite{jpcs24.1029,jlpt18.297}. Indeed, 
the gap is a decreasing function of increasing $\mu_\textrm{sc}$, and the largest value $|\Delta_0|$ occurs 
when $\mu_\textrm{sc}=0$. For $\mu_\textrm{sc}<0.60|\Delta_0|$ there is a second-order 
phase transition, while for $0.60|\Delta_0|<\mu_\textrm{sc}<0.707|\Delta_0|$ the 
gap $\Delta$ presents a discontinuous jump at the NM/SC transition temperature. 
For $\mu_\textrm{sc}>0.707|\Delta_0|$ the superconductivity is suppressed 
even at zero temperature. Here we considered only the scenario where 
$\mu_\textrm{sc}<0.60|\Delta_0|$.

The sd Hamiltonian accounts for a spin-flip process at the interface at which s-like electrons are
reflected, leading to the absorption (or emission) of angular momentum from (to) the magnetic side. 
Since we are considering a magnetic insulator, the interface interaction does not take into account conduction
electrons going into the magnetic side, and processes such as Andreev reflection are forbidden. However, the 
spin current injection across the interface is allowed due to the creation or annihilation of magnons (or 
spinon pairs of opposite spins). The interaction is expressed by

\begin{equation}
H_\textrm{sd}=J_\textrm{sd}\sum_{qkp}(S_q^- c_{k\uparrow}^\dagger c_{p\downarrow}+
S_q^+c_{p\downarrow}^\dagger c_{k\uparrow}),
\end{equation}
with $J_\textrm{sd}$ being a coupling constant. Here we consider
weak coupling between s-wave and d-wave electrons, so $H_\textrm{sd}$ is treated as a small
perturbation. In addition, a rough interface is assumed, which implies an independent transverse momentum sum. 

\section{Spin current}
We define the spin current operator as the time derivative of the difference
$N_\downarrow-N_\uparrow$ of electrons close to the interface. Using the Heisenberg equation, we obtain
$I=iJ_\textrm{sd}(V-V^\dagger)$, where the vertex operator is given by

\begin{equation}
V=\frac{1}{N}\sum_{qq^\prime kk^\prime}a_{q\downarrow}^\dagger a_{q^\prime\uparrow}c_{k\uparrow}^\dagger c_{k^\prime\downarrow}.
\end{equation}
Since we are considering the limit of weak interface interaction, the expected value 
$\langle I\rangle$ can be determined from the linear response theory. It is 
straightforward to obtain $I=\langle I\rangle=-i\hbar^{-1}\int dt\theta(t)\langle[\hat{I}(t),\hat{H}_\textrm{sd}(0)]\rangle$,
where the integral extends over the entire time axis and $\theta(t)$ denotes the Heaviside step
function. The caret denotes time evolution according
to $H_\textrm{m}+H_\textrm{sc}$, and since $N_\sigma=\sum_k c_{k\sigma}^\dagger c_{k\sigma}$ 
commutes with the full Hamiltonian $H=H_\textrm{m}+H_\textrm{sc}+H_\textrm{sd}$,
we can write $\hat{V}(t)=e^{i\Delta\mu t/\hbar}\tilde{V}(t)$, 
where the time evolution of $\tilde{V}$ is evaluated through the grand-canonical Hamiltonian. 
Therefore, we obtain

\begin{equation}
I=-\frac{2J_\textrm{sd}^2}{\hbar}\textrm{Im}U_\textrm{ret}(\Delta\mu),
\end{equation}
in which $U_\textrm{ret}(\Delta\mu)$ is the time Fourier transform of the 
retarded Green's function $\hbar U_\textrm{ret}(t)=-i\theta(t)\langle [\tilde{V}(t),\tilde{V}^\dagger(0)]\rangle$.
As usual, the retarded Green's function is determined by the Matsubara formalism, which 
provides $U_\textrm{ret}(\Delta\mu)$ through the analytical continuation of 
$\mathcal{U}(i\omega_l)=\int\mathcal{U}(\tau)e^{i\omega_l \tau}d\tau$, where

\begin{equation}
\label{eq.gtau}
\hbar \mathcal{U}(\tau)=-\langle T_\tau V(\tau)V^\dagger(0)\rangle=-\Xi_\textrm{m}(\tau) \Xi_\textrm{e}(\tau)
\end{equation}
is the imaginary-time Green's function. The magnetic term $\Xi_\textrm{m}$ of 
the Green's function is given by

\begin{equation}
\label{eq.xim}
\Xi_\textrm{m}(\tau)=\frac{1}{N^2}\sum_{qq^\prime}\mathcal{A}_{q\downarrow}(-\tau)\mathcal{A}_{q^\prime\uparrow}(\tau),
\end{equation} 
where we defined the a-operator Green's function 
$\mathcal{A}_{q\sigma}(\tau)=-\langle T_\tau a_{q\sigma}(\tau)a_{q\sigma}^\dagger(0)\rangle$.
Equation (\ref{eq.xim}) defines the annihilation of a $|q,\downarrow\rangle$ spinon state 
at the same time that a $|q^\prime,\uparrow\rangle$ state is created, resulting in 
an effective angular momentum variation of $\Delta S=\hbar$ in the magnetic insulator. Here
we have assumed dissipationless spin waves. However, if necessary, a damping term can easily be implemented
in the Green's function. On the other hand, the electronic part $\Xi_\textrm{e}$, 
written in terms of the b-operators, provides

\begin{figure}[h]
\centering \epsfig{file=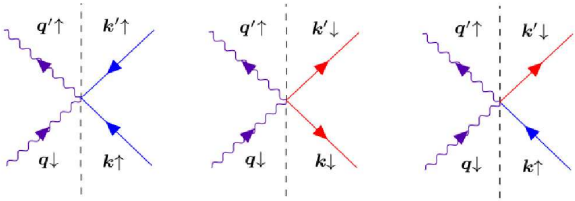,width=0.9\linewidth}
\caption{The three processes describing spin current injection at the 
interface: annihilation (left), creation (middle) and scattering (right) of quasiparticles
in the superconductor represented by the straight lines. The wavy lines represent
spinons in the magnetic insulator.}
\label{fig.interaction}
\end{figure}

\begin{IEEEeqnarray}{l}
\label{eq.xie}
\Xi_\textrm{e}(\tau)=\sum_{kk^\prime}\left[(|u_k v_{k^\prime}|^2-u_k v_k \bar{u}_{k^\prime}\bar{v}_{k^\prime})\mathcal{B}_{k\uparrow}(-\tau)\mathcal{B}_{k^\prime\uparrow}(-\tau)+\right.\nonumber\\
+(|u_k v_{k^\prime}|^2-u_k v_k \bar{u}_{k^\prime}\bar{v}_{k^\prime})\mathcal{B}_{k\downarrow}(\tau)\mathcal{B}_{k^\prime\downarrow}(\tau)+(u_k v_k\bar{u}_{k^\prime}\bar{v}_{k^\prime}+\nonumber\\
\left.+\bar{u}_k\bar{v}_k u_{k^\prime} v_{k^\prime}+|u_k u_{k^\prime}|^2+|v_k v_{k^\prime}|^2)\mathcal{B}_{k\uparrow}(-\tau)\mathcal{B}_{k^\prime\downarrow}(\tau)\right],
\end{IEEEeqnarray}
where $\mathcal{B}_{k\sigma}(\tau)=-\langle T_\tau b_{k\sigma}(\tau)b_{k\sigma}^\dagger(0)\rangle$ is
the Green's function associated with the b operators. The above equation describes
three different processes that decrease the spin on the SC side by $\hbar$, resulting in an
effective momentum angular transfer to the magnetic side. 
The first term in Eq. (\ref{eq.xie}) describes the annihilation of two up spin 
quasiparticles. Indeed, $\mathcal{B}_{k\uparrow}\mathcal{B}_{k^\prime\uparrow}$ is 
proportional to the occupation $f_{k\uparrow}f_{k^\prime\uparrow}$, where 
$f_{k\sigma}=f(E_{k\sigma})$ is the Fermi-Dirac distribution, 
while the multiplicative term in parentheses 
is the coherence factor. From Eq. (\ref{eq.b}), the $b_{k\uparrow}$ operator
gives a probability $|u_k|^2$ to annihilate an up-spin electron and
$|v_k|^2$ to create a down-spin electron. Therefore, the $|u_k v_{k^\prime}|^2$ 
term, for example, gives the probability of 
destroying a $|k,\uparrow\rangle$ electron state while a $|k^\prime,\downarrow\rangle$ 
electron state is created or, equivalently, a $|-k^\prime,\uparrow\rangle$ hole state is annihilated. 
Note that the charge is conserved in the process. In the same way, the second term represents the creation 
of two down spin quasiparticles, and the last one sets the scattering of an up spin to a 
down spin quasiparticle. All processes are represented in Fig. \ref{fig.interaction}. 

The spin current is then composed of the sum of three terms, $I=I^a+I^c+I^s$, where
the expected values $I^a$, $I^c$, and $I^s$ are the contributions associated 
with annihilation, creation, and scattering of quasiparticles at the interface, 
respectively. The analytical continuation $i\omega_l\to \Delta\mu+i0^+$ of the 
quasiparticle annihilation process $\mathcal{U}^a(i\omega_l)$, for example, provides

\begin{IEEEeqnarray}{l}
\textrm{Im} U_\textrm{ret}^a(\Delta\mu)=\frac{\pi\hbar}{4N^2}(e^{-\beta\Delta\mu}-1)\sum_{qq^\prime}(1+n_{q\downarrow})n_{q^\prime\uparrow}\sum_{kk^\prime}\left(1-\right.\nonumber\\
\left.-\frac{|\Delta|^2}{E_k E_{k^\prime}}\right)f_{k\uparrow}f_{k^\prime\uparrow}\delta(E_k+E_{k^\prime}+\hbar\omega_q-\hbar\omega_{q^\prime}).
\end{IEEEeqnarray}
In general, the energy scale of $\Delta\mu$ is much smaller than the thermal energy and
we adopted $1-e^{-\beta\Delta\mu}\approx \beta\Delta\mu$. 
After replacing the quasiparticle momentum sum by the continuum limit, we obtain

\begin{IEEEeqnarray}{l}
I^a(\Delta\mu)=\frac{\pi J_\textrm{sd}^2\beta\Delta\mu}{2N^2}\sum_{qq^\prime}n_{q\downarrow}(1+n_{q^\prime\uparrow})\int_0^\infty dE\int_{-\infty}^0 dE^\prime\left(1+\right.\nonumber\\
\left.+\frac{|\Delta|^2}{E E^\prime}\right)f(E_\uparrow)D(E)[1-f(E_\downarrow^\prime)]D(E^\prime)\delta(E-E^\prime+\hbar\omega_q-\nonumber\\
-\hbar\omega_{q^\prime}),
\end{IEEEeqnarray}
where $D(E)=\rho_F\textrm{Re}[(E+i\Gamma)/\sqrt{(E+i\Gamma)^2-|\Delta|^2}]$ is the 
superconducting density of states endowed by the phenomenological Dynes parameter $\Gamma$
and $\rho_F$ is the normal-metal density of states at the level Fermi. Note that $D(E)$ 
presents two narrow peaks at $E\approx \pm|\Delta|$ and tends to unity when $|E|\gg|\Delta|$ 
(the normal-metal limit). The inclusion of $\Gamma$ is necessary to ensure the 
convergence of the energy integral. To calculate the spin current, we adopted 
$\Gamma=0.05|\Delta_0|$ \cite{apl112.232601}. 
The $I^c$ and $I^s$ contributions are determined by the same procedure.

The magnetic part of the spin current requires special attention. In the ordered state,
the Schwinger boson condensation takes place, and the macroscopic population term $N_0$ 
needs to be removed from the momentum sum before we adopt the continuum limit. 
Although we are interested in the PMI/SC junction, the spin current 
can also be evaluated in other situations. Therefore, we write 
$n_{q^\prime\uparrow}=N_0\delta_{q,0}+n_{q^\prime\neq0\uparrow}$, where $N_0\approx N$
measures the condensation of up spinon states with $q=0$ (the limit of weak magnetic field 
$B\to 0^+$ is assumed). Summing over all quasiparticle processes and separating the condensate 
term from the $q^\prime$ sum, the spin current is written as $I=I_f+I_p$, where we define

\begin{IEEEeqnarray}{l}
\label{eq.If}
I_f(\Delta\mu)=\frac{N_0}{N}\frac{J_\textrm{sd}^2\beta\Delta\mu}{16\pi^2}\int_\textrm{BZ}d^3q n_{q\downarrow}\int_{-\infty}^\infty dE(1+\nonumber\\
\left. +\frac{|\Delta|^2}{E(E+\hbar\omega_q-\hbar\omega_0)}\right)f(E-\mu_\textrm{sc})D(E)[1-\nonumber\\
-f(E+\hbar\omega_q-\hbar\omega_0+\mu_\textrm{sc})]D(E+\hbar\omega_q-\hbar\omega_0)
\end{IEEEeqnarray}
as the ferromagnetic spin current associated with the up spinon condensation and

\begin{IEEEeqnarray}{l}
\label{eq.Ip}
I_p(\Delta\mu)=\frac{J_\textrm{sd}^2\beta\Delta\mu}{128\pi^5}\int_\textrm{BZ}d^3qd^3q^\prime n_{q\downarrow}(1+n_{q^\prime\uparrow})\int_{-\infty}^\infty dE(1+\nonumber\\
\left. +\frac{|\Delta|^2}{E(E+\hbar\omega_q-\hbar\omega_{q^\prime})}\right)f(E-\mu_\textrm{sc})D(E)[1-\nonumber\\
-f(E+\hbar\omega_q-\hbar\omega_{q^\prime}+\mu_\textrm{sc})]D(E+\hbar\omega_q-\hbar\omega_{q^\prime}).
\end{IEEEeqnarray}
as the paramagnetic spin current. In the above equations, the momentum integration is
done over the first Brillouin zone (BZ). Above the Curie temperature the condensation
vanishes ($N_0=0$), and the spin current is only due to the paramagnetic term. 
In the condensate phase, below the Curie transition temperature, $I_f$ shows an important role 
in the spin current behavior.

\begin{figure}[h]
\centering \epsfig{file=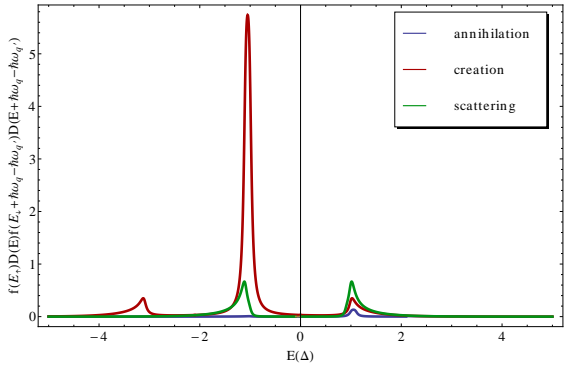,width=0.9\linewidth}
\caption{The integrand of the spin current energy integral
 for $\hbar\omega_q-\hbar\omega_{q^\prime}=0.1|\Delta|$ 
(quasiparticle scattering process), $\hbar\omega_q-\hbar\omega_{q^\prime}=2.1|\Delta|$
(quasiparticle creation process), and $\hbar\omega_q-\hbar\omega_{q^\prime}=-2.1|\Delta|$
(quasiparticle annihilation process). Here, $k_B T=0.5|\Delta|$, and $\mu_\textrm{sc}=0.01|\Delta|$.}
\label{fig.integrand}
\end{figure}

At zero temperature, the process of quasiparticle creation is the only relevant 
contribution to the spin current provided that 
$\hbar\omega_q-\hbar\omega_{q^\prime}>2|\Delta|$. At finite temperature, the largest 
spin current contributions occur when the peaks of $D(E)$ and 
$D(E+\hbar\omega_q-\hbar\omega_{q^\prime})$ coincide. There are three distinct cases:
(i) the quasiparticle scattering case, when $\hbar\omega_q-\hbar\omega_{q^\prime}\approx 0$,
(ii) the quasiparticle creation case for $\hbar\omega_q-\hbar\omega_{q^\prime}\approx 2|\Delta|$,
and (iii) the quasiparticle annihilation case when $\hbar\omega_q-\hbar\omega_{q^\prime}\approx-2|\Delta|$. The integrand of the spin current energy integral for $k_B T=0.5|\Delta|$ and $\mu_\textrm{sc}=0.01|\Delta|$ is shown
in Fig. \ref{fig.integrand}. The largest contribution occurs for the quasiparticle creation process when 
the highest peaks of $f(E_\uparrow)D(E)$ and 
$[1-f(E_\downarrow+\hbar\omega_q-\hbar\omega_{q^\prime})]D(E+\hbar\omega_q-\hbar\omega_{q^\prime})$ are close.

\section{Results}
\begin{figure}[h]
\centering \epsfig{file=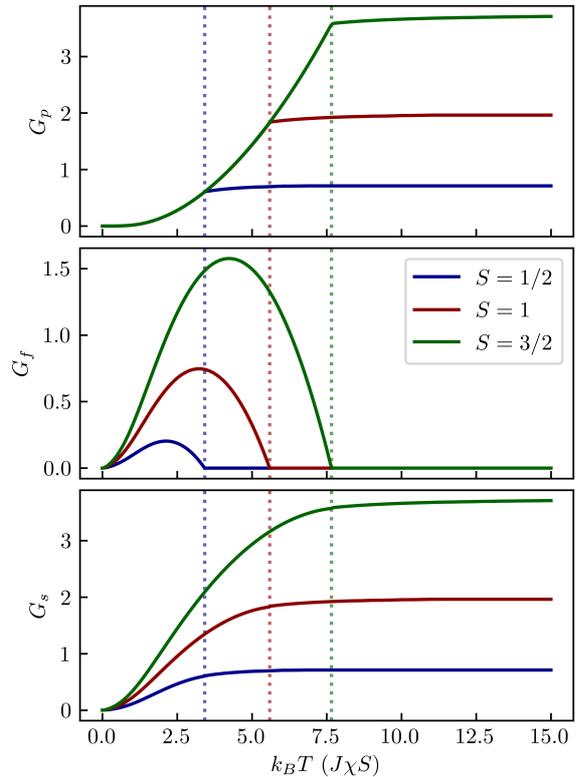,width=0.9\linewidth}
\caption{The paramagnetic (top) and ferromagnetic (center) contributions
 for the spin conductance for the normal-metal limit and the total spin conductance (bottom). 
 The vertical lines indicate the Curie transition temperature.}
\label{fig.i-nm}
\end{figure}

We are mainly interested in the spin current injection at the PMI/SC interface; however,
before we present the major results, to verify the model consistency we analyze 
other situations. As mentioned before, in the disordered phase, the ferromagnetic spin current $I_f$
vanishes due to the absence of spinon condensation; however, at very low temperatures,
$I_f$ has an important role. Recently, Okamoto \cite{prb93.064421} 
determined the spin current at the magnetic/normal-metal junction in both ordered 
and disordered phases using the Schwinger formalism. 
In the limit $T\to 0$, he found a spin current dependent on 
$T^3$ instead of the known result $I\propto T^{3/2}$ \cite{jpcs200.062030}. 
Okamoto associated the different power law behavior with the spin orientation of the
injected current. However, the inclusion of the condensate contribution 
$I_f$ restores the $T^{3/2}$ behavior. To see this, we consider the normal-metal
limit ($\Delta=0$) in the absence of magnetic field and the approximation

\begin{equation}
\int d\epsilon f(\xi_\uparrow)[1-f(\xi_\downarrow)]\approx \frac{(\hbar\omega_q+\Delta\mu) e^{\beta(\hbar\omega_q+\Delta\mu)}}{e^{\beta(\hbar\omega_q+\Delta\mu)}-1}.
\end{equation} 
Inserting the above result in Eq. (\ref{eq.If}), we obtain, for a small imbalance chemical potential,

\begin{IEEEeqnarray}{rCl}
I_f&=&\frac{N_0}{N}\frac{(\rho_F J_\textrm{sd})^2\beta\Delta\mu}{4\pi}\int_0^\infty dq \frac{\hbar\omega_q e^{\beta\hbar\omega_q}}{(e^{\beta\hbar\omega_q}-1)^2}\nonumber\\
&=&\frac{(\rho_F J_\textrm{sd})^2 S\beta\Delta\mu}{(J\chi S\pi)^{1/2}}(k_B T)^{3/2},
\end{IEEEeqnarray}
where the long-wavelength limit $\hbar\omega_q=J\chi S q^2$ was taken and we considered $N_0/N\approx 2S$. 
A similar procedure shows that $I_p\propto T^3$ due to the double Bose-Einstein distribution, and at 
low temperatures, we have $I\approx I_f\propto T^{3/2}$. Therefore, when the condensation term is
properly considered, we recover the expected power law dependence on $T$. The same result can be obtained 
from the Holstein-Primakoff formalism that is applicable to ordered states as well as the Schwinger
formalism in the condensate phase. In Fig. \ref{fig.i-nm} we show the spin conductance 
$G_\textrm{s}$ ($=\lim_{\Delta\mu\to 0}I/\Delta\mu$) associated with the ferromagnetic and paramagnetic
spin current contribution. The result obtained from the paramagnetic spin current is identical to that presented
in Ref. \cite{prb93.064421}; however, the ferromagnetic term gives a smoother transition to the
flat region above the Curie transition temperature. The plateau at high temperatures is provided by the
factor $k_B T$ that results from the energy integration and cancels the $\beta$ multiplicative factor and,
the paramagnetic boson condition $n_{q\uparrow}=n_{q\downarrow}$ that hinders the spinon scattering.  

\begin{figure}[h]
\centering \epsfig{file=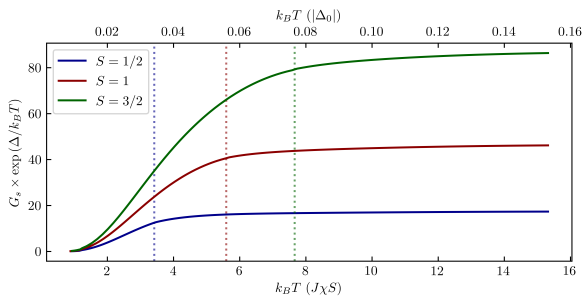,width=0.9\linewidth}
\caption{The spin current injection at the FM/SC interface. Due to the superconducting gap the
spin current intensity is multiplied by the Boltzmann factor $e^{-\beta|\Delta|}$ that causes 
a great reduction in $I$.}
\label{fig.i-fm-sc}
\end{figure}

Returning to the SC phase, we have two possible interfaces. The first one involves a FM/SC junction for 
which we adopt $J\chi\ll |\Delta_0|$. In this case, the spin current dependence on $T$ at very low temperatures 
(below the Curie transition temperature) is similar to that presented in paragraph above; however, 
the intensity is drastically reduced by a factor $e^{\beta|\Delta_0|}$. 
In the superconducting phase the probability that magnetic excitation has 
sufficient energy to induce spin injection is very low since $J\chi\ll |\Delta_0|$. 
Figure \ref{fig.i-fm-sc} shows the spin conductance behavior at low temperatures. In this limit, $I$ is
proportional to $e^{-\beta|\Delta_0|}T^{3/2}$, and when $|\Delta_0|\to 0$, we recover the result of
the FM/NM junction.

\begin{figure}[h]
\centering \epsfig{file=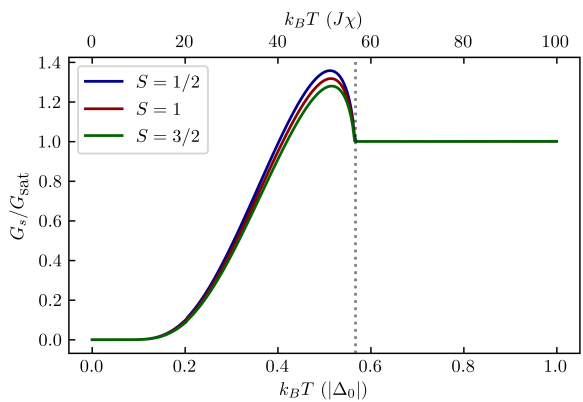,width=0.9\linewidth}
\caption{The spin current injection at the PMI/SC interface. The maximum above the SC transition temperature is provided by the 
coherence factor.}
\label{fig.g-pmi-sc}
\end{figure}

The second possible interface is the PMI/SC one. In this case we are considering temperatures 
on the interval $J\chi\lesssim k_B T\lesssim |\Delta_0|$ like in the GGG/NbN interface, for example. 
For $J\chi\ll|\Delta_0|$, the quasiparticle scattering process is the more relevant contribution to the spin current since 
spinons do not have sufficient energy to create or annihilate quasiparticles in the SC sample. 
In the paramagnetic phase, the spin current is given by Eq. (\ref{eq.Ip}) while $I_f=0$.
We choose $J\chi=0.01|\Delta_0|$, and the momentum integral of Eq. (\ref{eq.Ip}) is taken over the energy interval 
$|\hbar\omega_q-\hbar\omega_{q^\prime}|<0.02 |\Delta_0|$. The ratio $G_\textrm{s}/G_\textrm{sat}$ for 
$B=0$ as a function of the temperature is shown in Fig. \ref{fig.g-pmi-sc}. Here $G_\textrm{sat}$ stands for 
the NM spin conductance when the temperature tends to the SC transition point $T=0.568 |\Delta_0|^2/k_B$ from 
the values above. The spin conductance presents a peak below the SC transition temperature due to the 
coherence factor, while $G$ is equal to the NM spin conductance above the SC transition temperature. 
As one can note, above the SC transition point, the spin conductance is almost constant, and no visible 
variation is apparent. Our results provide the following peak values: 
$1.390 G_\textrm{sat}$ ($S=1/2$), $1.387 G_\textrm{sat}$ ($S=1$), and $1.384 G_\textrm{sat}$ ($S=3/2$). 
At very low temperatures, the magnetic ordered state occurs, and the spin conductance 
(as well as the spin current) is extremely small, as analyzed in the paragraph above. 

\begin{figure}[h]
\centering \epsfig{file=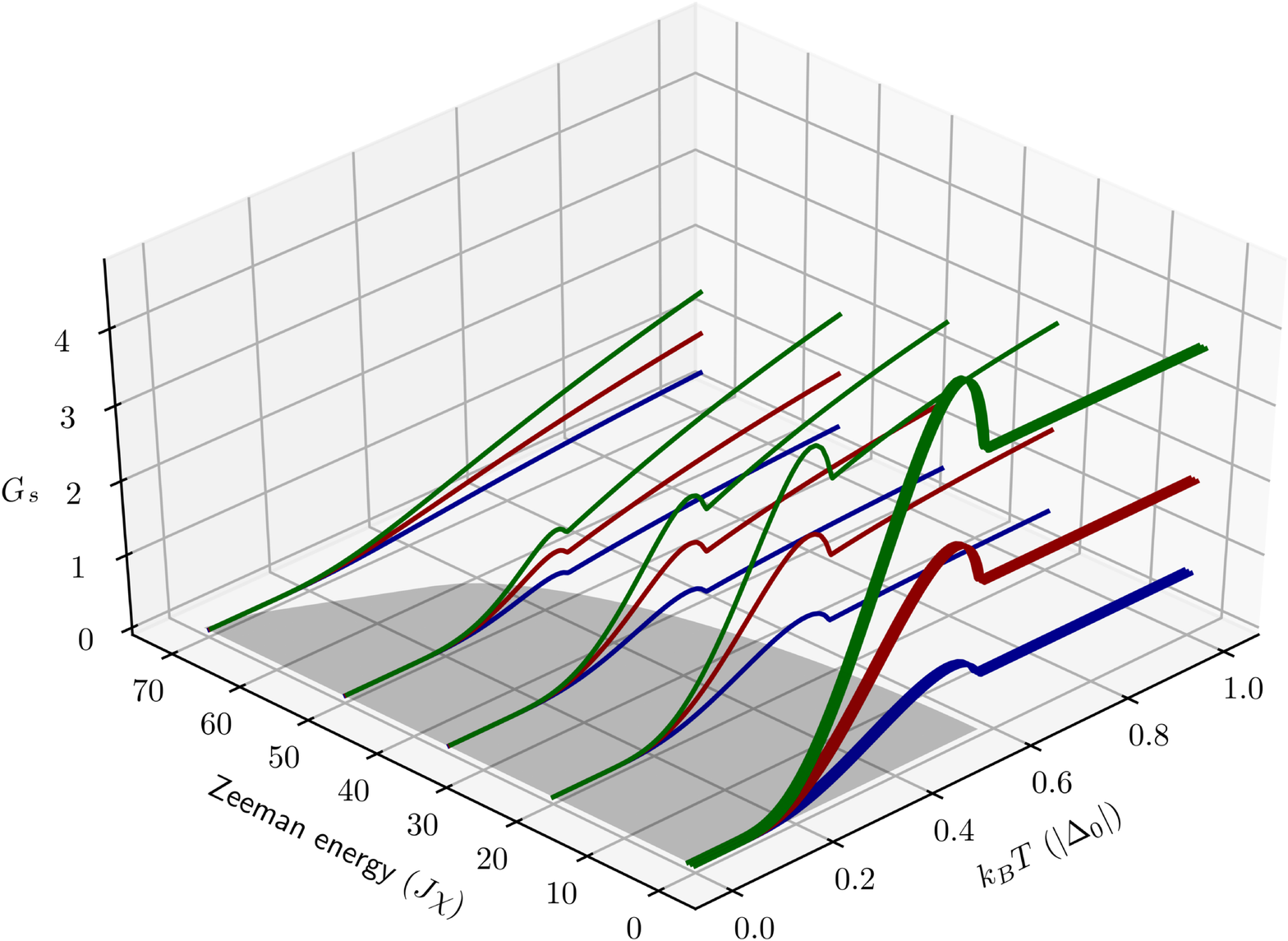,width=0.9\linewidth}
\caption{The spin conductance $G_s$ dependence on magnetic field (Zeeman energy) and
temperature. The effects of small magnetic field (with Zeeman energy of the order of $J\chi=0.01|\Delta_0|$) 
are negligible, while a strong magnetic field destroys the coherent behavior of the SC state. The superconductivity 
phase is represented by the shaded area. The blue, red, and green curves describe magnetic models
with spin $1/2$, 1, and $3/2$, respectively.}
\label{fig.mag_field}
\end{figure}

The magnetic field effect on spin conductance is shown in Fig. \ref{fig.mag_field}. The shaded area represents the 
superconductivity regime. Magnetic fields with Zeeman energy of the order of $J\chi$ have minimal effects on the 
spin conductance since we are adopting $J\chi =0.01|\Delta_0|$. The curves of magnetic field with energies of 0, 0.5, 
and 1 $J\chi$ present no visible difference. However, as is well known, high
magnetic fields suppress the superconductivity, and the quasiparticle coherence is destroyed.
The spin conductance for Zeeman energies larger than $0.707 |\Delta_0|$ (considering $\Delta\mu=0$) 
then shows the almost linear behavior $G_\textrm{s}\propto B$. 

\section{Summary and Conclusions}
In this work we investigated the spin current injection at the PMI/SC interface. The usual spintronics
experiments adopt junctions with ferromagnetic layers, and only recently has the role of disordered magnetic
been considered. Here we used the Schwinger formalism to treat both ordered and disordered magnetic phases
while the SC was described by the standard BCS theory. Therefore, we were able to identify two contributions
to the spin current. The first one (here denominated as ferromagnetic spin current) is associated with the 
condensate part of the Schwinger bosons, while the second one (called paramagnetic spin current) is due to the
bosons out of the condensate. In the limit of a vanishing SC gap, our equations provide the expected NM results
with minor corrections. In a recent work, for example, Okamoto \cite{prb93.064421} used the Schwinger 
formalism to evaluate the spin current at the PMI/NM interface, and he found a $T^3$ dependence at low 
temperature for the spin current instead of the expected $T^{3/2}$ behavior \cite{jpcs200.062030}. However, in 
that work, the condensate contribution was not taken into account. Meanwhile, in our results, the corrected 
$T^{3/2}$ power law of the spin current temperature dependence was obtained due to the condensate term. 
However, for a temperature above the Curie transition there is no boson condensation, and the 
paramagnetic spin current is the only relevant contribution. For the PMI/SC junction, the 
spin injection occurs mainly due to scattering of bogoliubons on the SC side, while the probability of 
quasiparticle creation (or annihilation) processes is very low since the exchange
coupling $J\ll|\Delta_0|$. Notwithstanding the lack of magnetic ordering, we showed an expressive 
spin current increasing at temperatures close to the SC transition. At $k_B T=0.511|\Delta_0$ the spin 
conductance shows an increase of approximately 40\% when compared to the NM value due to the 
coherence between quasiparticles in the SC state. In addition, low magnetic fields (of the order of 
$J\chi\sim 0.01|\Delta_0|$) present no perceptible effect on spin conductance, while high magnetic field 
(larger than the critical value of $0.707 |\Delta_0|/g\mu_B$) suppresses the superconductivity and
reduces the spin conductance.

This research was supported by CAPES (Finance Code 001).

\bibliography{manuscript}
\end{document}